\documentclass{rspublic}
\usepackage[sectionbib,square]{natbib}
\usepackage{graphicx}
\usepackage{amsmath, amssymb}
\usepackage{units}           

\newcommand{\gquote}[1]{`#1'} 
\newcommand{\avg}[1]{\langle #1 \rangle}
\newcommand{\twoCases}[4]{
  \left\{
    \begin{array}{ll}
      #1 & #2 \\
      #3 & #4
    \end{array}
  \right.
}

\begin{document}

\title[Impact of Adaptive Driving Strategies on Traffic Capacity]{Enhanced Intelligent Driver Model to Access the Impact of Driving Strategies on Traffic Capacity}

\author[A. Kesting, M. Treiber and D. Helbing]{Arne Kesting$^1$, Martin Treiber$^1$ and Dirk Helbing$^2$}
\affiliation{$^1$ Institute for Transport \& Economics, TU Dresden,\\
W\"urzburger Str. 35, 01187 Dresden, Germany\\
$^2$ ETH Zurich, UNO D11,\\
Universit\"atstr. 41, 8092 Zurich, Switzerland}

\label{firstpage}
\maketitle
\begin{abstract}{adaptive cruise control, human driving behaviour, car-following model, microscopic traffic simulation, free-flow capacity, dynamic capacity}
With an increasing number of vehicles equipped with adaptive cruise
control (ACC), the impact of such vehicles 
on the collective  dynamics of traffic flow becomes relevant. By means
of simulation, we investigate the influence of variable percentages of
ACC vehicles on traffic flow characteristics. For simulating the ACC
vehicles, we propose a new car-following model that also serves as
basis of an ACC implementation in real cars. The model is based on the
Intelligent Driver Model [Treiber \emph{et al.},  Physical Review E
62, 1805 (2000)] and inherits its intuitive behavioural parameters:
desired velocity, acceleration, comfortable deceleration, and desired
minimum time headway. It eliminates, however, the sometimes
unrealistic behaviour of the Intelligent Driver Model in cut-in
situations with ensuing small gaps that regularly are caused by lane
changes of other vehicles in dense or congested traffic. We simulate
the influence of different ACC strategies on the maximum  capacity before breakdown, and the (dynamic) bottleneck capacity after breakdown.  With a suitable strategy, we find sensitivities of the order of 0.3, i.e., 1\% more ACC vehicles will lead to an increase of the capacities by about 0.3\%. This sensitivity multiplies when considering  travel times at actual breakdowns. 
\end{abstract}

\section{\label{sec:intro}Introduction}
Efficient transportation systems are essential to the functioning and
prosperity of modern, industrialized societies. Engineers are
therefore seeking solutions to the questions of how the capacity of
the road network could be used more efficiently and how operations can
be improved by way of intelligent transportation systems
(ITS). Achieving this efficiency through automated vehicle control is
the long-standing vision in transport telematics. With the recent
advent of advanced driver assistance systems, at least partly
automated driving is already available for basic driving tasks such as
accelerating and braking by means of \emph{adaptive cruise control}
(ACC) systems. An ACC system extends earlier cruise control to
situations with significant traffic in which driving at constant speed
is not possible. The driver cannot only adjust the desired velocity
but also set a certain safe time gap determining the distance to the
leading car when following slower vehicles. 

Although originally developed to delineate human driving behaviour, \emph{car-follow\-ing models} can also be used to describe ACC systems: A radar sensor tracks the car ahead to measure the net distance (gap) and the approaching rate, which serve (besides the own speed) as input quantities just as in many time-continuous car-following models. Then, the ACC system calculates the appropriate acceleration for adapting the speed and the safety gap to the leader. This analogy is scientifically interesting because a \gquote{good} car-following model could serve as basis of a control algorithm of a real-world ACC system. On the one hand, the in-vehicle implementation would allow to judge the realism of the considered car-following model which is a further (and promising) approach towards benchmarking the plethora of car-following models. This question has recently been considered with different methods in the literature~\citep{Brockfeld-benchmark04,Ossen-benchmark05,Punzo-bench05,Panwai-Calibration-2005,Kesting-Calibration-TRR08}. On the other hand, from the direct driving experience one may expect new insights for the (better) description of human drivers through adequate follow-the-leader models, which is still an ongoing challenge in traffic science~\citep{HDM,Ossen-interDriver06,Kerner-book,Helb-Opus}.

Furthermore, one may raise the interesting question how the \emph{collective traffic dynamics} will be influenced in the future by an increasing number of vehicles equipped with ACC systems. Microscopic traffic simulations are the appropriate methodology for this since this approach allows to treat  \gquote{vehicle-driver units} individually and in interaction. The results, however, may significantly depend on the chosen modelling assumptions. In the literature, positive as well as negative effects of ACC systems have been reported~\citep{Arem_ACC_Impact_Transaction,Arne-ACC-TRC,Davis-ACC,VanderWerf-TRR-2002,marsden-ACC}. This puzzling fact points to the difficulty when investigating mixed traffic consisting of human drivers and automatically controlled vehicles: How to describe human and automated driving and their interaction appropriately?

The Intelligent Driver Model (IDM)~\citep{Opus} appears to be a good
basis for the development of an ACC system. The IDM shows a crash-free
collective dynamics, exhibits controllable stability
properties~\citep{Helb-Phases-EPJB-09}, and implements an intelligent
braking strategy with smooth transitions between acceleration and
deceleration behaviour. Moreover, it has only six parameters with a
concrete meaning which makes them measurable. However, the IDM was
originally developed as a simple car-following model for one-lane
situations. Due to lane changes (\gquote{cut-in} manoeuvres), the
input quantities change in a non-continuous way, in which the new
distance to the leader can drop significantly
below the current equilibrium distance, particularly if there is 
dense or congested traffic, the lane change is mandatory, or if 
drivers have different conceptions of safe distance.
This may lead to strong braking manoeuvres of the IDM, which would not be acceptable
(nor possible) in a real-world ACC system. 

In this paper, we therefore extend the IDM by a new
\emph{constant-acceleration heuristic}, which implements a more relaxed
reaction to cut-in manoeuvres without loosing the
mandatory model property of being essentially crash-free. This model
extension has already been implemented (with some further confidential
extensions) in real test cars~\citep{Kranke-Fisita2008,Kranke-VDI-2006-engl}. In a second part
of this contribution we apply the enhanced IDM to multi-lane traffic
simulations in which we study the collective dynamics of mixed traffic
flows consisting of human drivers and adaptive cruise control
systems. The vehicles equipped with ACC systems implement a recently
proposed traffic-adaptive driving strategy~\citep{Arne-ACC-TRC}, which
is realized by a situation-dependent parameter setting for each
vehicle.

Our paper is structured as follows: In \S\ref{sec:model}, we will present an improved heuristic of the IDM particularly suited for multi-lane simulations. Section \S\ref{sec:vla} presents traffic-adaptive driving strategies for adaptive cruise control systems. The impact of temporarily changed model parameters on the relevant traffic capacities in heterogeneous traffic flows will be systematically evaluated by simulations in \S\ref{sec:sim}. Finally, we will conclude with a discussion in \S\ref{sec:disc}.

\section{\label{sec:model}A Model for ACC Vehicles}
In this section, we will develop the model equations of the
enhanced Intelligent Driver Model. To this, we will first present  
the relevant aspects of the 
original Intelligent Driver Model (IDM)~\citep{Opus}
in~\S\ref{sec:model}$\,$(\ref{sec:IDM}). While, in most
situations, the IDM
describes accelerations and decelerations in a satisfactory way, 
it can lead to unrealistic driving behaviour if the actual vehicle gap is significantly
lower than the desired gap, and, simultaneously, the situation can be
considered as only mildly critical.

Therefore, in~\S\ref{sec:model}$\,$(\ref{sec:CAH}),
we develop an upper limit of a safe acceleration  based on the
more optimistic \gquote{constant-acceleration heuristic} (CAH), 
in which drivers assume that
the leading vehicle will not change its acceleration for the next few
seconds. This ansatz is applicable precisely in these situations where
the IDM reacts too strongly. In~\S\ref{sec:model}$\,$(\ref{sec:IDM-CAH-mix}), we combine the IDM and CAH
accelerations to specify the acceleration function
of the final model for ACC vehicles (\gquote{ACC model}) 
such that the well-tested IDM is applied whenever it leads to a
plausible behaviour, using the difference $a_\text{CAH}-a_\text{IDM}$
as an indicator for plausibility. Finally, the properties of the
new model are tested by computer simulations in~\S\ref{sec:model}$\,$(\ref{sec:CAH-sim}).

\subsection{\label{sec:IDM}Intelligent Driver Model}
The IDM acceleration is a continuous function incorporating different
driving modes for all velocities in freeway traffic as well as city
traffic. Besides the (bumper-to-bumper-) distance $s$ to the leading
vehicle and the
actual speed $v$, the IDM also takes into account the velocity difference
(approaching rate) $\Delta v=v - v_\text{l}$ to the leading vehicle.
The IDM acceleration function is given by
\begin{eqnarray}
\label{eq:IDMaccel}
a_\text{IDM}(s,v,\Delta v)  &=& \frac{\text{d}v}{\text{d} t} =  a
         \left[ 1 -\left( \frac{v}{v_0} \right)^\delta -\left(
         \frac{s^*(v,\Delta v)} {s}
         \right)^2\, \right], \\
\label{eq:IDMsstar}
s^*(v, \Delta v)  &=& s_0  + v T + \frac{v \Delta v }  {2\sqrt{a b}}.
\end{eqnarray}
This expression combines the free-road acceleration strategy
$\dot{v}_\text{free} (v)= a[1-(v/v_0)^\delta]$ with a deceleration strategy
$\dot{v}_\text{brake}(s, v, \Delta v) = -a(s^*/s)^2$ 
that becomes relevant when the gap to the leading vehicle is not
significantly larger than the effective \gquote{desired (safe) gap}
$s^*(v, \Delta v)$.
The free acceleration is characterized by the
\textit{desired speed} $v_0$, the
\textit{maximum acceleration} $a$, and the exponent $\delta$
characterizing how the acceleration decreases  with velocity
($\delta=1$ corresponds to a linear decrease while $\delta\to \infty$
denotes a constant acceleration). The effective minimum gap $s^*$
is composed of the \textit{minimum distance} $s_0$ (which is relevant
for low velocities only), the velocity dependent distance $vT$ which corresponds
to following the leading vehicle with a constant \emph{desired time
gap} $T$, and  a dynamic contribution which is only active in non-stationary traffic
corresponding to situations in which $\Delta v\ne 0$. This latter contribution implements an
\gquote{intelligent} driving behaviour that, in normal situations,
 limits braking decelerations to the
\emph{comfortable deceleration} $b$. In critical situations,
however, the IDM deceleration  becomes significantly higher,
making the IDM \emph{collision-free}~\citep{Opus}. 
The IDM parameters
$v_0$, $T$, $s_0$, $a$ and $b$ (see table~\ref{tab:params})
have a reasonable interpretation, are known to be relevant,
are empirically measurable, and have realistic
values~\citep{Kesting-Calibration-TRR08}.

\begin{table}
\centering

\caption{\label{tab:params}Model Parameters}
\longcaption{Parameters of the Intelligent Driver Model and the ACC
model used in the simulations of \S\ref{sec:model}$\,$(\ref{sec:sim}). The first six
parameters are common for both models. The last parameter is
applicable to the ACC model only.}
\begin{tabular}{lcc}
\hline
Parameter                       & Car & Truck \\ \hline
Desired speed $v_0$             & \unit[120]{km/h} & \unit[85]{km/h} \\
Free acceleration exponent $\delta$     & 4 & 4 \\
Desired time gap $T$            & \unit[1.5]{s}    & \unit[2.0]{s}\\
Jam distance $s_0$              & \unit[2.0]{m}    & \unit[4.0]{m} \\
Maximum acceleration $a$        & \unit[1.4]{m/s$^2$} & \unit[0.7]{m/s$^2$} \\
Desired deceleration $b$        & \unit[2.0]{m/s$^2$} & \unit[2.0]{m/s$^2$} \\
Coolness factor $c$             & 0.99 & 0.99 \\
\hline
\end{tabular}
\end{table}

\subsection{\label{sec:CAH}Constant-Acceleration Heuristic}

The braking term of the IDM is developed such that accidents are
avoided even in the \emph{worst case}, where the driver of the
leading vehicle suddenly brakes with the maximum possible 
deceleration $b_\text{max} \gg b$ to a complete
standstill. Since the IDM does not include explicit reaction
times, it is even safe when the  time headway parameter $T$ is set to
zero.\footnote{Notice that reaction time and time headway are conceptionally
different quantities, although they have the same order of
magnitude. Optionally, an effective reaction time 
can be implemented by a suitable update time in the numerical integration of the 
IDM acceleration equation~\citep{ThreeTimes-07}.}

However, there are situations, characterized by comparatively low
velocity differences and gaps that are significantly smaller than the
desired gaps, where this \textit{worst-case
heuristic} leads to overreactions. In fact, human drivers simply
rely on the fact that the drivers of preceding vehicles will not
suddenly initiate full-stop emergency brakings without any reason and,
therefore, consider such
situations only as mildly critical. Normally this judgement is
correct. Otherwise,  the frequent observations of accident-free driving
at time headways significantly below \unit[1]{s}, i.e., below the
reaction time of even an attentive driver, 
would not be observed so frequently~\citep{HDM}. 

In order to characterize this more optimistic view of the drivers, let us
investigate the implications of the 
\emph{constant-acceleration heuristic} (CAH) on the safe
acceleration. The CAH is based on the following assumptions:
\begin{itemize}
\item The accelerations of the considered and leading vehicle will not
change in the relevant future (generally, a few seconds).
\item No safe time headway or minimum distance is required at any
moment.
\item Drivers react without delay (zero reaction time).
\end{itemize}
When calculating the maximum acceleration for which the situation
remains crash-free, one needs to distinguish whether the velocity of
the leading vehicle is zero or nonzero at the time where the minimum gap
(i.e., $s=0$) is reached. 
For given actual values of the gap $s$, velocity $v$, velocity
$v_l$ of the leading vehicle, and its acceleration $a_l$, the maximum
acceleration $a_\text{CAH}$ leading to no crashes is given by

\begin{equation}
\label{accCAH}
a_\text{CAH}(s,v,v_l, a_l) = \twoCases{
 \frac{v^2 \tilde{a}_l}{v_l^2 - 2 s \tilde{a}_l}}
 {\text{if} \ \  v_l (v-v_l)  \le  -2 s \tilde{a}_l,}
 {\tilde{a}_l - \frac{(v-v_l)^2\Theta(v-v_l)}{2s}} {\text{otherwise,}}
\end{equation}
where the effective acceleration $\tilde{a}_l=\min(a_l,a)$ has been used to avoid artefacts that may be 
caused by leading vehicles with higher acceleration capabilities.
The condition $v_l (v-v_l) = v_l \Delta v \le -2 s a_l$ is true if
the vehicles have stopped at the time the minimum gap $s=0$ is reached.
Otherwise, negative approaching rates do not make sense to the CAH and
are therefore eliminated by the Heaviside step function $\Theta$.

\begin{figure}
\centering
\includegraphics[width=\textwidth]{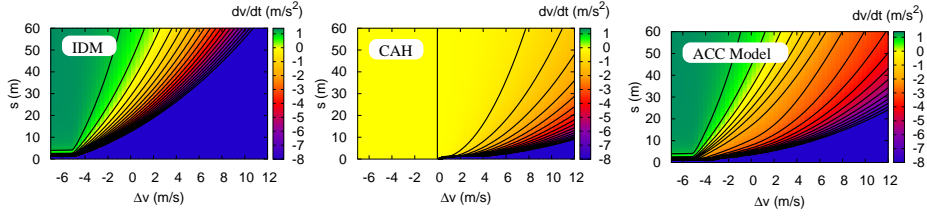}

\caption{\label{fig:accfunctions-vleadfix}Acceleration functions of (a)~the IDM,
(b)~resulting from the CAH heuristic, (c)~of the proposed ACC model as a
function of the gap $s$ and the velocity difference (approaching rate)
$\Delta v$ to the leading vehicle. The velocity of the leading vehicle and its acceleration $a_l$ are given by
$v_l=\unit[20]{m/s}$ and $a_l=0$, respectively. }
\end{figure}

In figure~\ref{fig:accfunctions-vleadfix}(b), the CAH acceleration~\eqref{accCAH} has been plotted for a leading vehicle driving at
constant velocity. A comparison with 
figure~\ref{fig:accfunctions-vleadfix}(a) clearly shows that, for small
values of the gap $s$, the CAH
acceleration is significantly higher (i.e., less negative) than that
for the IDM.

\subsection{\label{sec:IDM-CAH-mix}The ACC Model}

For the situations where the IDM leads to
unnecessarily strong braking reactions, the CAH acceleration is
significantly higher, corresponding to a more relaxed reaction. 
This heuristic, however, fails on the other side and therefore is
not suited to directly model the accelerations of ACC vehicles.
Specifically, the CAH leads to 
zero deceleration for some cases that clearly require at
least a
moderate braking reaction. This includes a
stationary car-following situation ($\Delta v=0$, $a_\text{l}=0$), where
$a_\text{CAH}=0$ for arbitrary
values of the gap $s$ and velocity $v$, see figure~\ref{fig:accfunctions-vleadfix}(b).
Moreover, since the CAH does not include minimum time headways or
an acceleration to a
desired velocity, it does not result in a  complete model.

\begin{figure}
\centering
\includegraphics[width=0.55\textwidth]{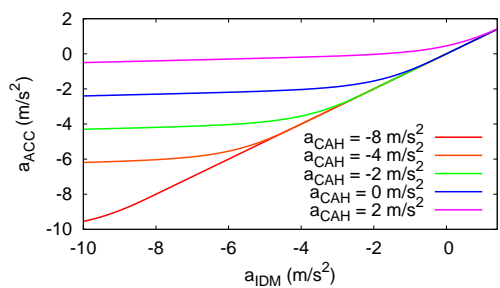}

\caption{\label{fig:accMix}Resulting acceleration $a_\text{ACC}$ of the ACC model, equation~\protect\eqref{accACC}, as a function of the IDM acceleration for different values of the  acceleration $a_\text{CAH}$ resulting from the constant-acceleration heuristic (CAH).}
\end{figure}

For the actual formulation of a model for ACC vehicles, we will
therefore use the CAH only as an indicator to  determine whether 
the IDM will lead to unrealistically high decelerations, or
not. Specifically, the proposed ACC model is based on following assumptions:

\begin{itemize}
\item  The  ACC acceleration is never lower than that of the IDM. This is motivated by the circumstance that the IDM will lead  to crash-free vehicle trajectories for all simulated situations.
\item If both, the IDM and the CAH, produce the same acceleration,  the 
ACC acceleration is the same as well.
\item If the IDM produces extreme decelerations, while the CAH
yields accelerations greater than  $-b$, the situation is considered
as mildly critical and the ACC acceleration is
essentially equal to the comfortable deceleration plus a
small fraction $1-c$ of the IDM deceleration. 
\item If both the IDM and the CAH result in accelerations
significantly below $-b$, the situation is seriously critical and the
ACC acceleration must not be higher than the maximum of the IDM and
CAH accelerations. 
\item The ACC acceleration should be a continuous and differentiable function of the IDM
and CAH accelerations.
\end{itemize}
Probably the most simple functional form satisfying these criteria is
given by (cf. figure~\ref{fig:accMix})
\begin{equation}
\label{accACC}
a_\text{ACC} =\twoCases{a_\text{IDM}}{a_\text{IDM}\ge a_\text{CAH},}
{(1-c)\, a_\text{IDM} + c \left[
a_\text{CAH} + b \tanh \left(\frac{a_\text{IDM}-a_\text{CAH}}{b}\right)\right]}
{\text{otherwise.}}
\end{equation}
This acceleration equation of the ACC model is the main model-related result of this
paper. Figure~\ref{fig:accfunctions-vleadfix} shows that the
conditions listed above are fulfilled. Notably, the ACC model leads to
more relaxed reactions in situations in which the IDM behaves
too conservatively. 
In contrast to the IDM, the acceleration depends not only on
the gap to and the velocity of the leading vehicle, but (through $a_\text{CAH}$) on the
\textit{acceleration} $a_l$ of this vehicle as well.
This leads to a more defensive driving behaviour when approaching 
congested traffic (reaction to \gquote{braking lights}), but also to a more
relaxed behaviour in  typical cut-in situations where a slower vehicle
changes to the fast lane (e.g., in order to overtake a truck) while
another vehicle in the fast lane is approaching from behind.

Compared with the IDM parameters, the ACC model contains only one
additional model parameter $c$, which can be interpreted as a
\textit{coolness factor}. For $c=0$, the model reverts to the IDM,
while for $c=1$ the sensitivity with respect to changes of the gap
vanishes in situations with small gaps and no velocity difference.
This means that the behaviour would be too relaxed. Realistic values for the coolness
factor are in the range $c\in [0.95, 1.00]$. Here, we have assumed $c=0.99$,
cf. table~\ref{tab:params}.

\subsection{\label{sec:CAH-sim}Simulating the Properties of the ACC Model}

\begin{figure}
\centering
\includegraphics[width=\textwidth]{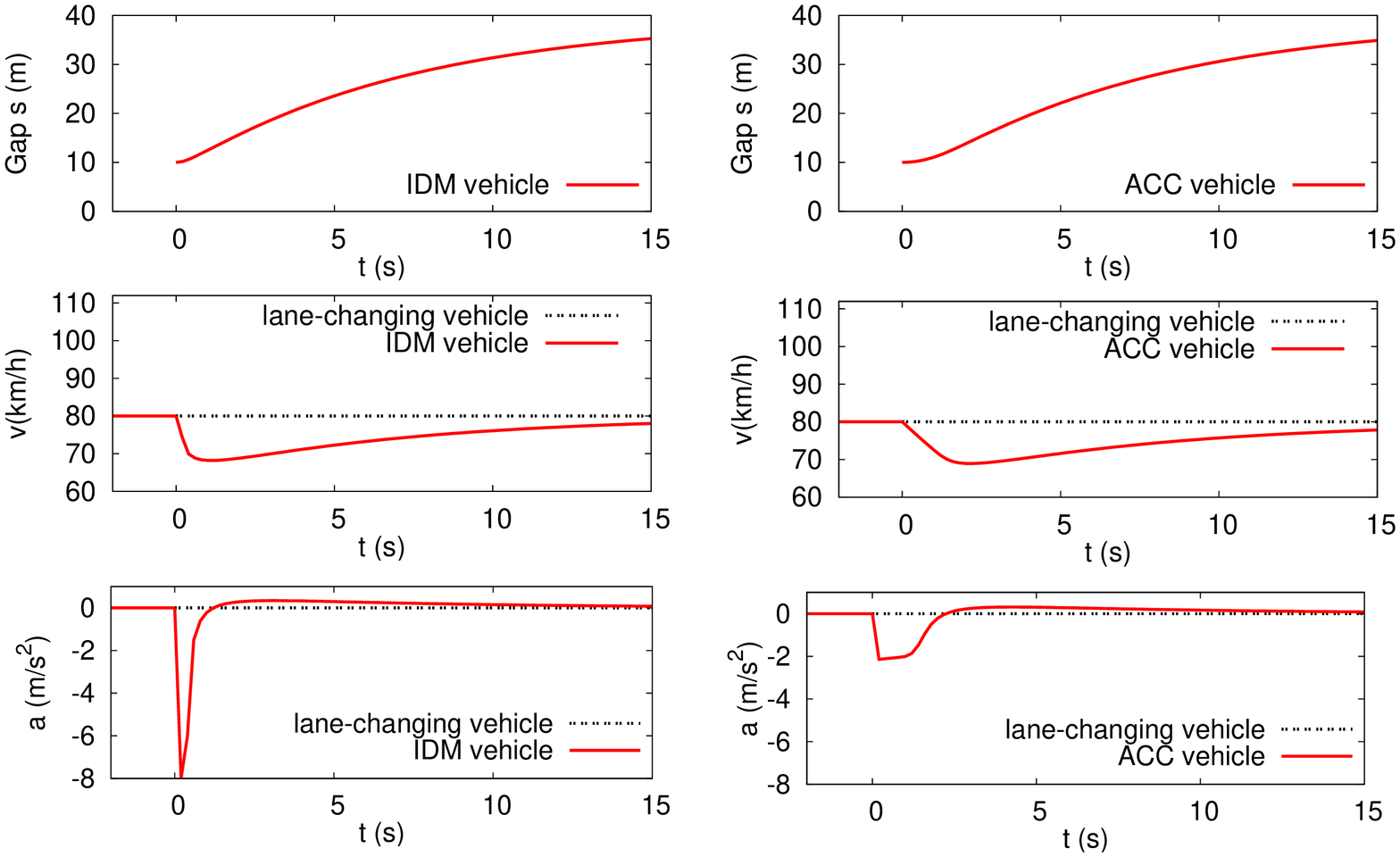}

\caption{\label{fig:cutin_dv0}Response of an IDM and an ACC vehicle
(\gquote{car} parameters of table~\ref{tab:params}) 
to the lane-changing manoeuvre of another vehicle immediately in front of
the considered vehicle. The initial velocities of both vehicles is
\unit[80]{km/h}, and the initial gap is \unit[10]{m}. This can be
considered as a \gquote{mildly critical} situation.
}
\end{figure}

Figure~\ref{fig:cutin_dv0} displays simulations of a mildly critical
\gquote{cut-in} situation: Another vehicle driving at the same
velocity \unit[80]{km/h} merges at time $t=0$ in front of the
ACC vehicle (car parameters of table~\ref{tab:params}) 
resulting in an initial gap $s(0)=\unit[10]{m}$. 
Although the corresponding time
headway $s(0)/v(0)= \unit[0.45]{s}$ is less than one third of the
desired minimum time headway 
$T=\unit[1.5]{s}$, the
situation is not really critical because the approaching rate at the
time of the lane change is zero (the time-to-collision $s/\Delta v$ is
even infinite). Consequently, the ACC model  results in a braking deceleration
that does not exceed the comfortable deceleration $b=\unit[2]{m/s^2}$.
In contrast,  the IDM leads to decelerations reaching 
temporarily the maximum value  assumed to be
physically possible ($\unit[8]{m/s^2}$). In spite of the more relaxed
reaction, the velocity drop of the ACC vehicle is slightly less than that of the IDM
(minimum velocities of about \unit[69]{km/h} and \unit[68]{km/h}, respectively).

\begin{figure}
\centering
\includegraphics[width=\textwidth]{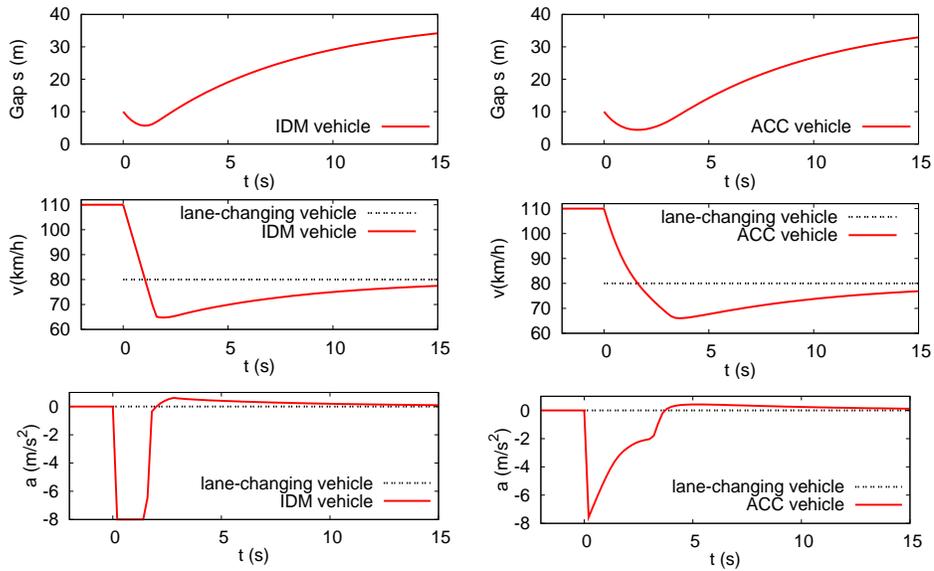}

\caption{\label{fig:cutin_dvLarge}Response of an IDM and an ACC vehicle
to an abrupt lane-changing manoeuvre of another vehicle immediately in front of
the considered vehicle. The decelerations are
restricted to $\unit[8]{m/s^2}$. The initial velocity of the lane-changing
vehicle is \unit[80]{km/h}, while the initial velocity of the 
considered vehicle is \unit[110]{km/h}. This is a \gquote{strongly critical} situation. }

\end{figure}

A seriously critical situation is depicted in
figure~\ref{fig:cutin_dvLarge}. While the \gquote{cut-in} results in the
same initial gap $s(0)=\unit[10]{m}$ as in the situation discussed
above, the ACC vehicle is approaching rapidly (initial approaching
rate $\Delta v(0)=\unit[30]{km/h}$) and an emergency braking is
mandatory to avoid a rear-end crash. For this case, the  reactions of
the two models are
similar. Both the IDM and the ACC models lead to initial decelerations
near the maximum value. However, the deceleration of the ACC vehicle quickly
decreases towards the comfortable deceleration before reverting to
a stationary situation. In contrast, the IDM vehicle remains in emergency mode for
nearly the
whole braking manoeuvre. Of course, the  more relaxed ACC behaviour leads
to a closer minimum gap  (\unit[4]{m} compared to \unit[5.5]{m} for
the IDM). Nevertheless, the velocity drop of the ACC vehicle is
slightly lower than that of the IDM (minimum velocities of about~\unit[66]{km/h} compared to~\unit[64]{km/h}).

\begin{figure}
\centering
\includegraphics[width=0.55\textwidth]{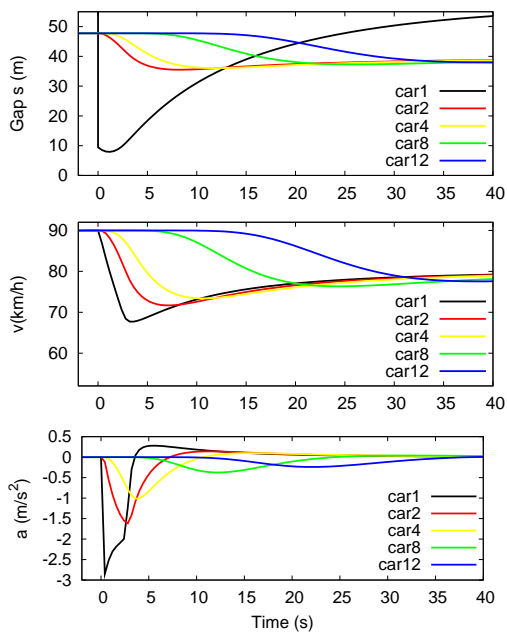}

\caption{\label{fig:cutin_platoon}Response of a platoon of ACC vehicles
to an abrupt lane-changing manoeuvre of another vehicle
($v=\unit[80]{km/h}$) in front of the platoon 
with an initial gap of \unit[10]{m} only. The desired velocity of the
platoon leader is \unit[90]{km/h}, while that of the followers is
\unit[120]{km/h}. 
}
\end{figure}

Both the maximum deceleration and
the velocity drop are measures for the perturbations imposed on
following vehicles and therefore influence the stability of a platoon of
vehicles, i.e., the \gquote{string stability}.
While softer braking reactions generally lead to a reduced string
stability, the reduced perturbations behind ACC vehicles generally
compensate for this effect.
In fact, figure~\ref{fig:cutin_platoon} shows for a specific example, that the
excellent stability properties of the IDM carry over to the ACC model.
Further simulations show that, for the \gquote{car} model parameter
set of table~\ref{tab:params} and also for heterogeneous traffic
composed of up to 40\% trucks, string stability is satisfied for nearly
all situations. An analytical investigation to delineate the stability
limits is the topic of  future work.

\section{\label{sec:vla}Driving Strategy for Adaptive Cruise Control Systems}

In this section we summarize an extension of adaptive cruise control (ACC) systems towards traffic-condition-dependent driving strategies, which has recently been proposed by the authors~\citep{Arne-ACC-TRC}. Since drivers require to have full control of and confidence in their car, an ACC system should be designed in a way that its driving characteristics is similar to the natural driving style of human drivers. In order to nevertheless improve traffic performance by automated driving, we therefore propose a \emph{traffic-adaptive driving strategy} which can be implemented by changing the model parameters depending on the local traffic situation. 

For an efficient driving behaviour it is sufficient to change the default driving behaviour only temporarily in specific traffic situations. The situations in which the parameters are dynamically adapted have to be determined autonomously by the equipped vehicle~\citep{Arne-ACC-TRC}. To this end, we consider the following discrete set of five traffic conditions that are characterized as follows:
\begin{enumerate}
\item \emph{Free traffic}. This is the default situation. The
ACC settings are determined solely by the maximum individual driving
comfort. Since each driver can set his or her own parameters for the
time gap and the desired speed, this may lead to different
settings of the ACC systems.
\item \emph{Upstream jam front}. Here, the objective is to increase safety by reducing 
velocity gradients. Compared to the default situation, this implies
earlier braking when approaching slow vehicles. Note that the
operational layer always assures a safe approaching process
independent of the detected traffic state.
\item \emph{Congested traffic}. Since drivers cannot influence
the development of traffic congestion in the bulk of a traffic jam,  
the ACC settings are reverted to their default values.
\item \emph{Downstream jam front}. To increase the dynamic bottleneck capacity,
accelerations are increased and time gaps are temporarily decreased.
\item \emph{Bottleneck sections}. Here, the objective is to locally
increase the capacity, i.e., to \emph{dynamically bridge the capacity
gap}. This requires a temporary reduction of the time gap. Moreover, by increasing the
maximum acceleration, the string stability of a vehicle platoon is increased due
to a shorter adaptation time to changes in the velocities~\citep{ThreeTimes-07}.
\end{enumerate}

Each traffic condition can be associated with certain settings for the model parameters implementing the intended driving style. For the enhanced Intelligent Driver Model (cf.\ \S\ref{sec:model}), the relevant parameters are the desired time gap $T$, the desired maximum acceleration $a$, and the desired deceleration $b$. In order to preserve the individual settings of the drivers or users, respectively, changes in the parameter sets for each traffic state can be formulated in terms of multiplication factors. These relative factors $\lambda$ can be arranged in a \emph{\gquote{driving strategy matrix}} shown in table~\ref{tab:strategymatrix}. For example, $\lambda_T=0.7$ denotes a reduction of the default time gap~$T$ by 30\% in bottleneck situations.

\begin{table}
\caption{\label{tab:strategymatrix}Summary of the ACC driving strategies}

 \longcaption{Each of the traffic conditions corresponds to a different
 set of ACC control parameters. The ACC driving characteristics are represented by the IDM parameters \emph{ safety time gap}~$T$, the \emph{maximum
 acceleration}~$a$ and the \emph{comfortable deceleration}~$b$. $\lambda_T$, $\lambda_a$, and $\lambda_b$ are multiplication factors. }

\centering
\begin{tabular}{lcccl}
\hline
Traffic condition & $\quad\lambda_T\quad$ & $\quad\lambda_a\quad$ & $\quad\lambda_b\quad$  &  Driving behaviour \\\hline
Free traffic      & 1 & 1 & 1      & Default/Comfort \\
Upstream front    & 1 & 1 & 0.7    & Increased safety  \\
Congested traffic & 1 & 1 & 1      & Default/Comfort \\
Downstream front  & 0.5 & 2 & 1    & High dynamic capacity \\
Bottleneck        & 0.7 & 1.5 & 1  & Breakdown prevention \\\hline
\end{tabular}
\end{table}

As indicated in the strategy matrix, the default settings (corresponding to $\lambda=1$) are used most of the time. For improving traffic performance, only two traffic states are relevant. The regime \gquote{passing a
bottleneck section} aims at an
suppression (or, at least, at a delay) of the traffic flow collapse by
lowering the time gap $T$ in combination with an increased
acceleration $a$.\footnote{Note that a
local reduction in the road capacity is the \emph{defining
characteristics} of a bottleneck. Consequently, the proposed driving
style in the bottleneck section should lead to a \emph{dynamic
homogenization} of the road capacity, thereby allowing for a higher
maximum flow at the bottleneck.} Furthermore, a short-term reduction of traffic
congestion can only be obtained  by increasing the outflow.  This is the
goal in the traffic condition \gquote{downstream jam front}, aiming at a brisk leaving of the queue by increasing the maximum acceleration $a$ and decreasing the
safe time gap $T$ of the ACC system. Both regimes are systematically evaluated in the following section.

\section{\label{sec:sim}Impact of the ACC Driving Strategies on Capacities}

In this section we investigate the impact of ACC-equipped vehicles implementing the traffic-adaptive driving strategy on the traffic performance by systematically varying the  given ACC proportion. The philosophy of our simulation approach is to control the system by keeping as many parameters constant (\emph{ceteris paribus} methodology). Consequently, we start with a well defined system (with 0\% ACC) and vary the ACC fraction as external parameter. In free flow, the \emph{maximum  throughput} of a freeway is determined by the maximum flow occurring before the
traffic flow breaks down, while in congested traffic it is given by
the \emph{dynamic capacity} (i.e., the outflow from a
traffic jam). These capacities will be studied in
\S\ref{sec:sim}$\,$(\ref{sec:max_free}) and \S\ref{sec:sim}$\,$(\ref{sec:dyn_capa}). 

\subsection{\label{sec:max_free}Maximum Flow in Free Traffic}

The relevant measure for assessing the efficiency of the proposed
driving strategy \gquote{passing a bottleneck section} is the maximum
possible flow until the traffic flow breaks down. An upper bound for
this quantity $C$, defined as the maximum number of vehicles per unit time
and lane capable of passing the bottleneck, is given by the inverse of the safe time gap $T$, i.e.,
$C<1/T$. However, the theoretical
maximum flow also depends on the effective length  $l_\text{eff}=l+s_0$ of a driver-vehicle
unit  (which is given by the vehicle length $l$
plus the minimum bumper-to-bumper distance $s_0$). Therefore, the theoretically possible maximum flow is 
\begin{equation}\label{eq:qmax}
Q_\text{max}^\text{theo} = \frac{1}{T} \left(1 - \frac{l_\text{eff}}{v_0 T+l_\text{eff}} \right).
\end{equation}
This \emph{static road capacity} $Q_\text{max}^\text{theo}$
corresponds to the maximum of a \emph{triangular} fundamental
diagram. In the case of the IDM (with a finite acceleration), the
theoretical maximum is even lower~\citep{Opus}. Generally, the \emph{maximum free flow}
$Q_\text{max}^\text{free}$ before traffic breaks down is a \emph{dynamic} quantity that depends on the traffic stability as
well. Typically, we have $Q_\text{max}^\text{free} \le Q_\text{max}^\text{theo}$. 

In the following, we will therefore analyse the maximum free flow
$Q_\text{max}^\text{free}$ resulting from traffic simulations. To
this end, we consider a simulation scenario with a two-lane freeway and an
on-ramp which creates a bottleneck situation. The lane-changing decisions have been modelled using MOBIL~\citep{MOBIL-TRR07}. The inflow at the upstream boundary
was increased with a constant rate of $\dot{Q}=\unit[700]{veh/h^2}$, while
the ramp flow was kept constant at \unit[250]{veh/h/lane}. We have
checked other progression rates as well, but only found a marginal
difference. In order to determine the maximum free flow, we have used the
following criterion: A traffic breakdown is detected, if more than 20~vehicles on the main road drive slower than $v_\text{crit}=\unit[30]{km/h}$. After a traffic breakdown has been detected, we
use the flow of the actual 1-min aggregate of a \gquote{virtual} detector
located downstream of the on-ramp to  measure the maximum flow.

\subsubsection{Probability of a breakdown of traffic flow}

The maximum free flow $Q_\text{max}^\text{free}$ results from a
measurement process which is based on 1-min aggregation intervals. As
the underlying complex traffic simulation involves nonlinear models,
discrete lane change decisions, random influences (such as the
vehicle type inserted at the upstream
boundary etc.), it is expected that $Q_\text{max}^\text{free}$ is a stochastically varying quantity, leading to different measurements even for identical
boundary and initial conditions (assuming a random seed of
the computer's pseudo-random number generator). Consequently, we will consider the
maximum free flow as a \emph{random variable}
which reflects the probabilistic nature of a traffic flow breakdown~\citep{Geistefeldt-stochCap-TRR,Kerner-book,persaud-breakdown1998}.

\begin{figure}
\centering

\begin{tabular}{cc}
\includegraphics[width=0.6\textwidth]{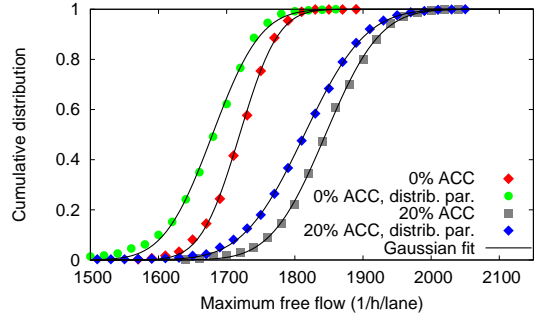}
\end{tabular}

 \caption{\label{fig:ACC_breakdownProb_truck}
Traffic breakdown probability for an ACC equipment rate of 0\% and 20\%, respectively, and for different degrees of heterogeneity. The diagrams show the cumulative distribution functions of the maximum free flow, resulting from 1000 simulation runs and numerical fits for the Gaussian distribution. }
\end{figure}

The statistical properties of $Q_\text{max}^\text{free}$ have been
investigated by means of repeated simulation runs. We have examined scenarios with a total truck percentage of 10\% with an additional ACC equipment level of 20\% and without ACC-equipped vehicles. Furthermore, we have considered \emph{inter-driver variability} by assigning uniformly and independently distributed
values to the IDM parameters $v_0$, $T$, $a$ and $b$, where the averages of the parameter values have been left unchanged, while the width of the distributions have been set to 20\% (i.e., the
individual values vary between 80\% and 120\% of the average parameter
value). Each scenario has been simulated 1000 times to derive the statistical
properties of the maximum free flow. The resulting cumulative
distribution functions for $Q_\text{max}^\text{free}$, reflecting the probability of a traffic
flow breakdown, are shown in figure~\ref{fig:ACC_breakdownProb_truck}. As the measurement of the maximum free flow yields a distribution of
finite variance and results from many stochastic contributions, the
resulting cumulative distribution function is expected to follow the \emph{Central
Limit Theorem}. In fact, the integrated (and normalized) Gaussian function $N(x; \mu, \sigma^2)$
 with mean value $\mu$ and variance $\sigma^2$ fits the simulation results well.

From the results in figure~\ref{fig:ACC_breakdownProb_truck}, we draw the following conclusions: 
First, an increased proportion of ACC vehicles
shifts the maximum throughput to a larger mean value. This shows the
positive impact of the traffic-adaptive driving strategy on the
traffic efficiency. In particular, the temporary change of the driving
characteristics while passing the bottleneck helps to increase the
maximum throughput by 6--8\% in the considered scenarios (at an ACC
equipment level of $\alpha=20\%$). 

Second, the considered traffic scenarios show different degrees of
heterogeneity, i.e., mixtures of various vehicle types (cars and
trucks, equipped with ACC system or not) and statistically distributed model parameters. An increase in
the degree of heterogeneity leads to larger fluctuations in the
traffic flow which, in turn, results in a larger variation $\sigma$ of
the random variable $Q_\text{max}^\text{free}$ and also in a slightly
reduced mean value. 

Third, as the consideration of ACC-equipped vehicles with their adaptive
(i.e., \emph{time-dependent}) parameter choice increases the level of
heterogeneity in a significant way, the variation $\sigma$ is
increased compared to the values without ACC vehicles. This finding
makes clear that the impact of ACC-equipped vehicles on the traffic
dynamics must be studied with a realistic level of heterogeneity,
e.g., in a multi-lane freeway scenario considering cars and
trucks. Otherwise, the assessment of ACC systems may erroneously lead
to discouraging (or overly optimistic) results.

\subsubsection{Maximum free flow as a function of the ACC proportion}

\begin{figure}
\centering

  \begin{tabular}{cc}
 \includegraphics[width=0.45\textwidth]{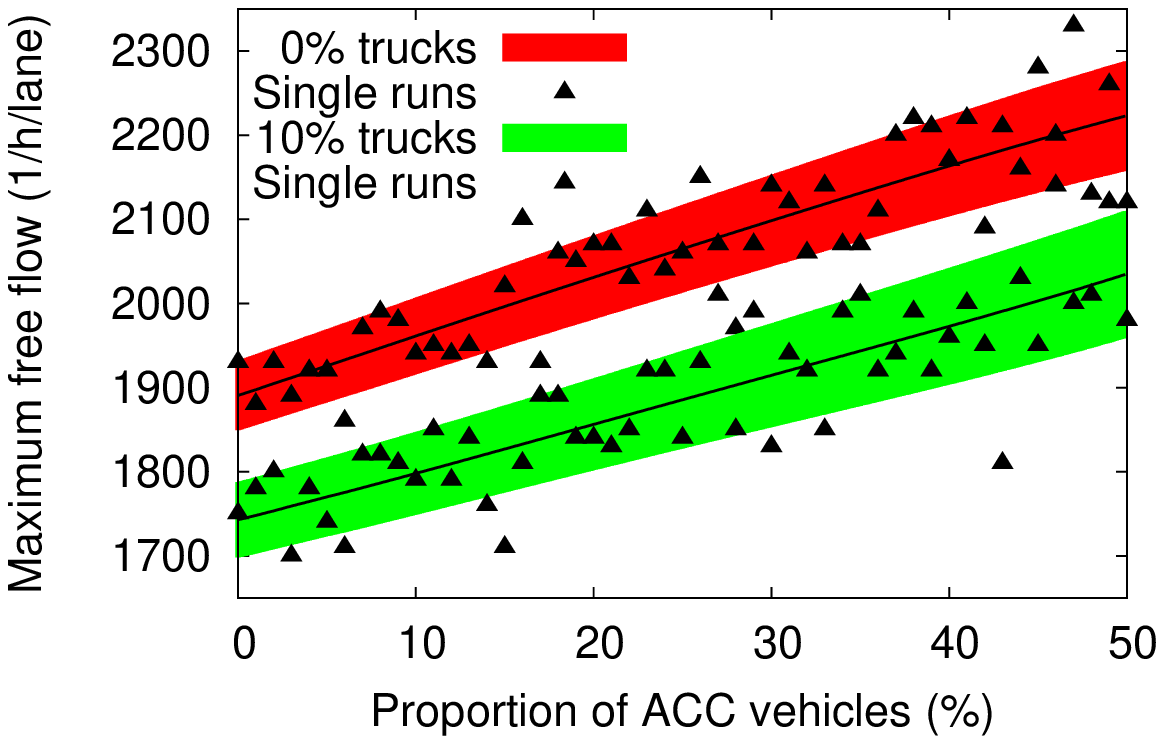} & 
 \includegraphics[width=0.45\textwidth]{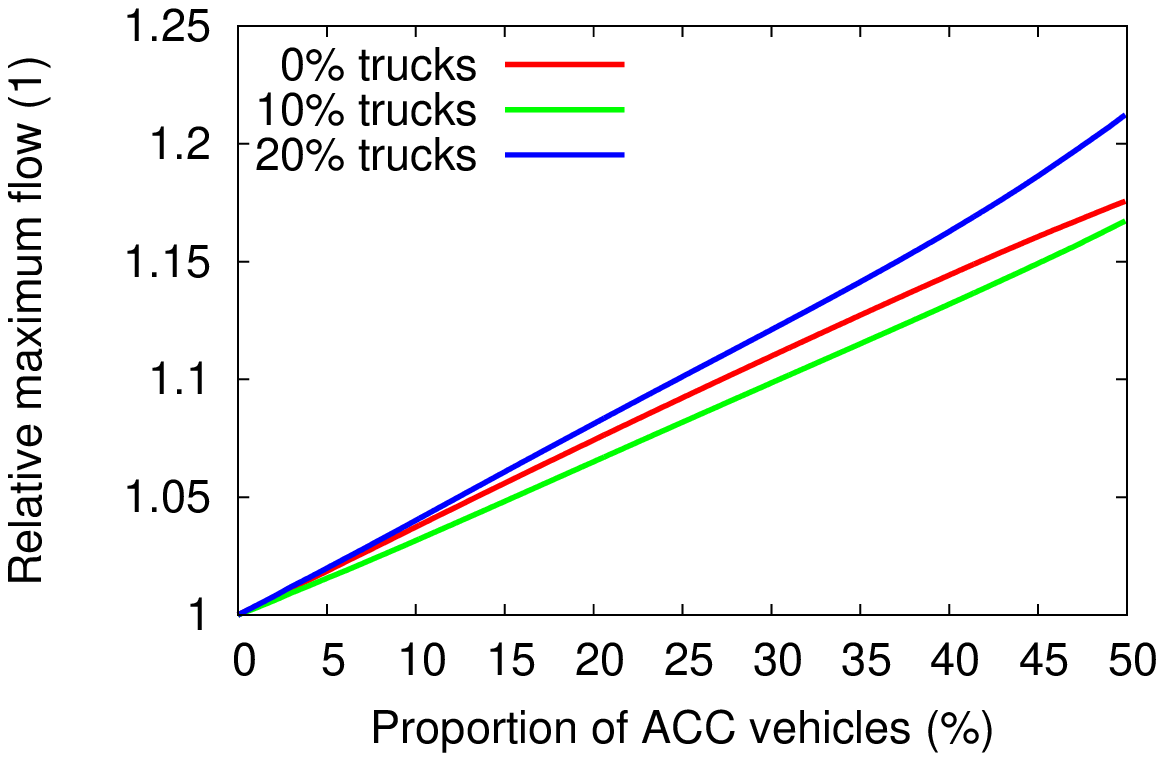}
\end{tabular}

 \caption{\label{fig:free_vla}Maximum free flow until
 traffic breaks down as a function of the ACC proportion for various
 truck fractions. Note that the genuine cumulative distributions for $\alpha=0\%$
 and $\alpha=20\%$ are shown in figure~\ref{fig:ACC_breakdownProb_truck}.}
\end{figure}

Let us now investigate the maximum free flow as a function of the ACC
proportion $\alpha$. To this end, we have gradually increased $\alpha$
in a range from $0$ to $50\%$, using the scenario described
above. In figure~\ref{fig:free_vla}, the result of each simulation run
is represented by one point. As expected from our previous findings, the
values of the maximum free flow $Q_\text{max}^\text{free}$ vary
stochastically. For better illustration, we have therefore used a
\emph{kernel-based linear regression}, which calculates the
expectation value and the standard deviation as a \emph{continuous}
function of $\alpha$. The only parameter of this evaluation procedure is the width $\delta$
of the smoothing kernel. Here, we have used $\delta=0.1$. Details of the
numerical method are described in~\ref{app:regression}.

Figure~\ref{fig:free_vla}(a) shows the results for a traffic scenario
without trucks and a scenario with $10\%$ trucks. As expected,
increasing the proportion of trucks with their higher safe time gap
$T$ (cf.\ table~\ref{tab:params}) reduces the maximum free
flow. However, the average value of the maximum free flow increases
with growing  ACC equipment level $\alpha$. \emph{The gain in the maximum
free flow is basically proportional to $\alpha$.}

Figure~\ref{fig:free_vla}(b) summarizes the simulation results for
various truck proportions in terms of the \emph{relative increase} $q_\text{max}^\text{free}= Q_\text{max}^\text{free}(\alpha) / Q_\text{max}^\text{free}(0)$ of the maximum flow compared to situations with non-equipped vehicles. This quantity allows for a direct comparison between the different
simulation scenarios. For example, the gain in the maximum free flow
varies between approximately 16\% and 21\% for an ACC fraction of
50\%. For an ACC portion of 20\%, the maximum free flow increases
by approximately 7\%. Although this appears to be a relatively small enhancement, one
should not underestimate its impact on the resulting traffic
dynamics. The authors, for example, have shown that 
an ACC proportion of 20\% can often prevent (or, at least, delay) a breakdown of traffic
flow~\citep{Arne-ACC-TRC}. Comparing this to the reference simulation without
\gquote{intelligent} ACC-equipped vehicles, individual travel times vary by a
factor of~2 or~3 at least, sometimes even more. As the gain in the maximum free flow is basically
proportional to $\alpha$, the quantity
$q_\text{max}^\text{free}/\alpha$ is approximately constant and
describes the \emph{relative gain} in $Q_\text{max}^\text{free}$ per
ACC portion~$\alpha$. The values for the simulation results shown in
figure~\ref{fig:free_vla} are in the range between $0.32$ and $0.42$.

\subsubsection{Maximum flow for different driving strategy parameters}

\begin{figure}
\centering
\begin{tabular}{cc}
 \includegraphics[width=0.45\textwidth]{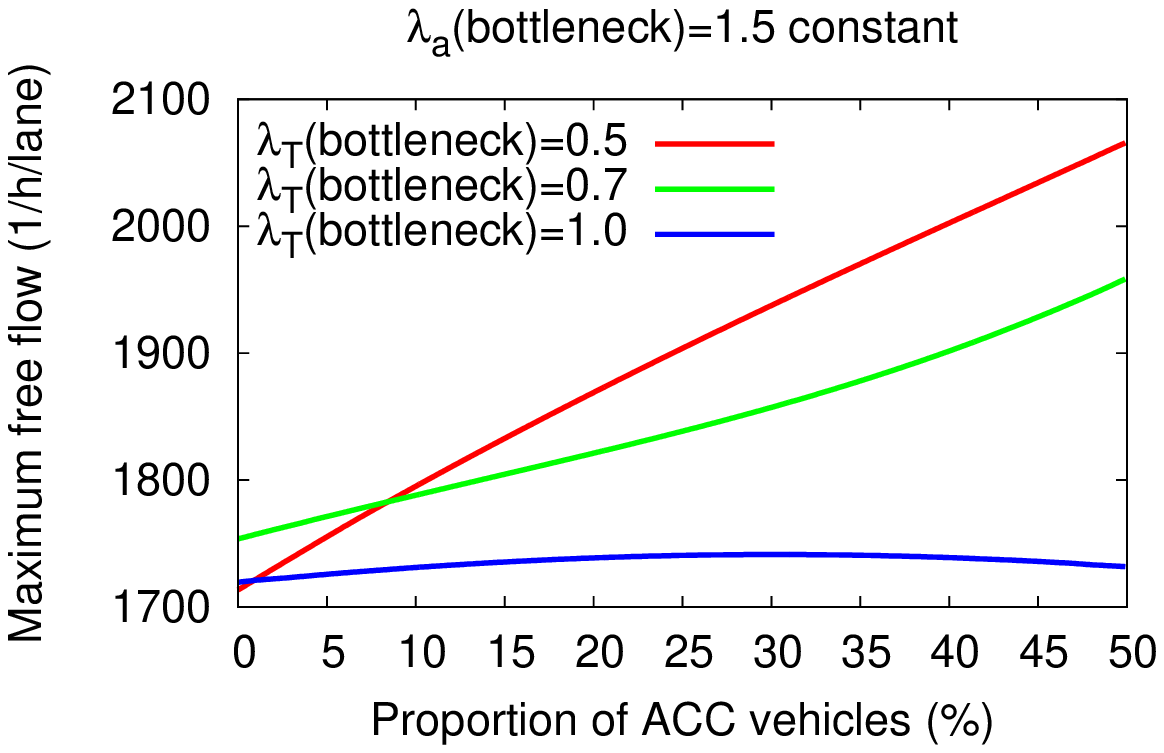} & 
 \includegraphics[width=0.45\textwidth]{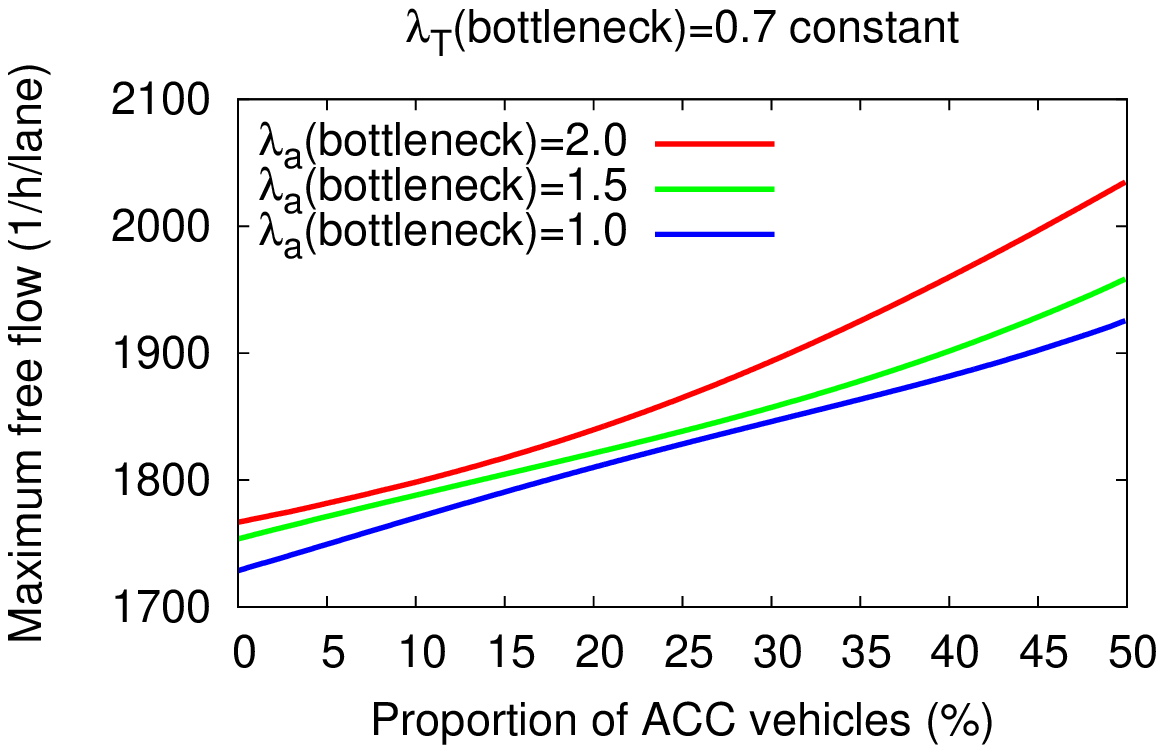}
\end{tabular}
 \caption{\label{fig:ACC_free_matrix}Average maximum free flow as a function of
 the ACC proportion for the simulation scenario with 10\% trucks. The
 ACC driving strategy in the \gquote{bottleneck} state is varied: The left
 diagram shows various settings for $\lambda_T$ while keeping the
 maximum acceleration $a$ constant. The right diagram refers to
 simulation results for various values of $\lambda_a$, while the safe time gap $T$ is kept constant.}

\end{figure}

Besides the proportions of trucks and ACC-equipped vehicles, the
traffic performance is influenced by the multiplication factors
$\lambda$ of the ACC driving strategy in the bottleneck state. In
particular, the maximum free flow depends on the modification
$\lambda_T$ of the time gap and $\lambda_a$ of the maximum
acceleration in the \gquote{bottleneck condition}. As default values, we have
chosen $\lambda_T^\text{bottle}=0.7$ and $\lambda_a^\text{bottle}=1.5$
 (see table~\ref{tab:strategymatrix}). While considering the
aforementioned simulation scenario with 10\% trucks, we have varied
these driving strategy parameters in the \gquote{bottleneck} state as shown in figure~\ref{fig:ACC_free_matrix}.

The left diagram in figure~\ref{fig:ACC_free_matrix} shows the strong
impact of the safe time gap $T$ on the maximum free flow (while
keeping $\lambda_a^\text{bottle}=1.5$ constant). A further reduction 
 of the ACC time gap with $\lambda_T=0.5$ leads to a stronger increase
 in the maximum free flow when considering a growing ACC
proportion. Note that this is 
 consistent with equation~\eqref{eq:qmax}. Moreover, the modification of
 $a$ alone while keeping $T$ unchanged ($\lambda_T=1$) does not
 improve the maximum free flow. 

The maximum acceleration $a$ has clearly a smaller effect on
 the maximum free flow than $T$, as displayed in the right diagram of
 figure~\ref{fig:ACC_free_matrix}. While keeping $\lambda_T$ constant,
 different settings such as $\lambda_a=1$, $1.5$ or $2$ do not change the
 resulting maximum free flow in a relevant way. So, the throughput can
only be efficiently increased \emph{in combination with smaller time
gaps} (corresponding to lower values of $\lambda_T$).

\subsection{\label{sec:dyn_capa}Dynamic Capacity after a Traffic Breakdown}
Let us now investigate the system dynamics \emph{after} a breakdown of
traffic flow. Traffic jam formation is determined by the \emph{difference} of the upstream inflow $Q_\text{in}$ and the outflow $Q_\text{out}$ from the downstream jam front
(i.e., the head of the queue) which is also called \emph{dynamic
capacity}. We use the same simulation setup (of a two-lane freeway with an on-ramp)
as in the previous section. Whenever a traffic breakdown was 
provoked by the increasing inflow, we aggregated the flow data of
the \gquote{virtual detector} $1\,$km downstream of the bottleneck within an
interval of 10 minutes.

\subsubsection{Dynamic capacity as a function of the ACC proportion}
\begin{figure}
\centering

\begin{tabular}{ccc}
\includegraphics[width=0.45\textwidth]{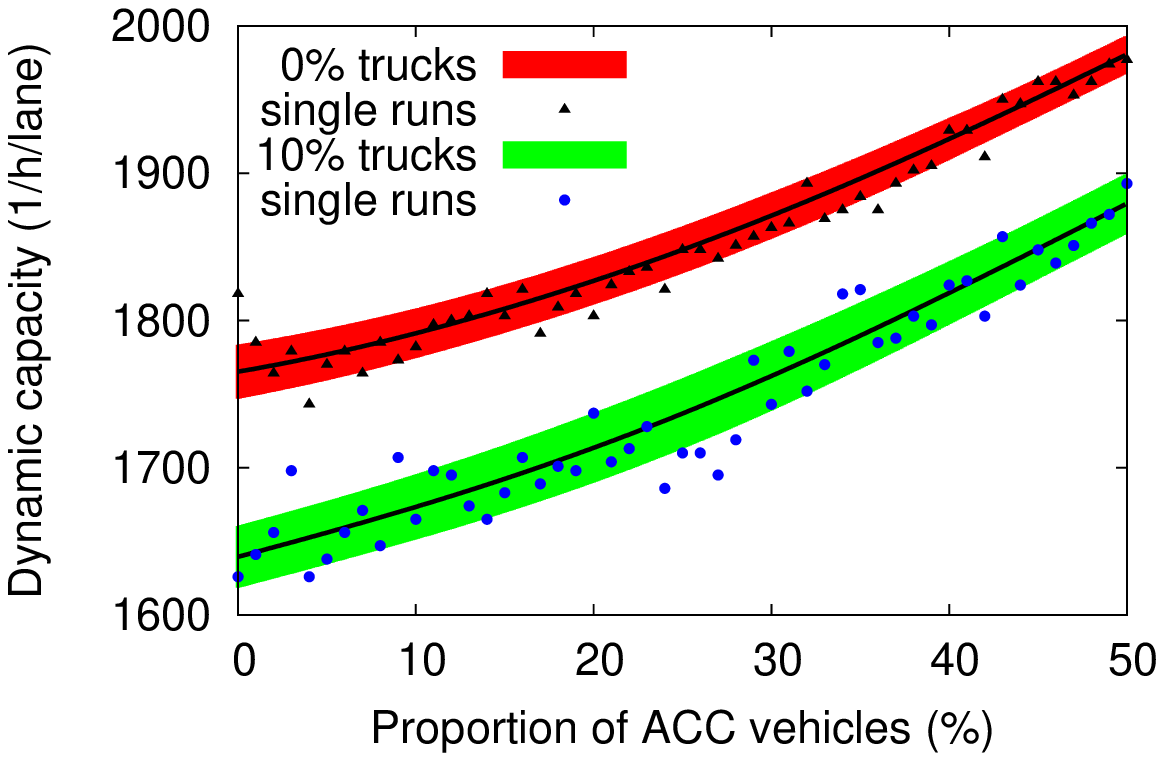} &
\includegraphics[width=0.45\textwidth]{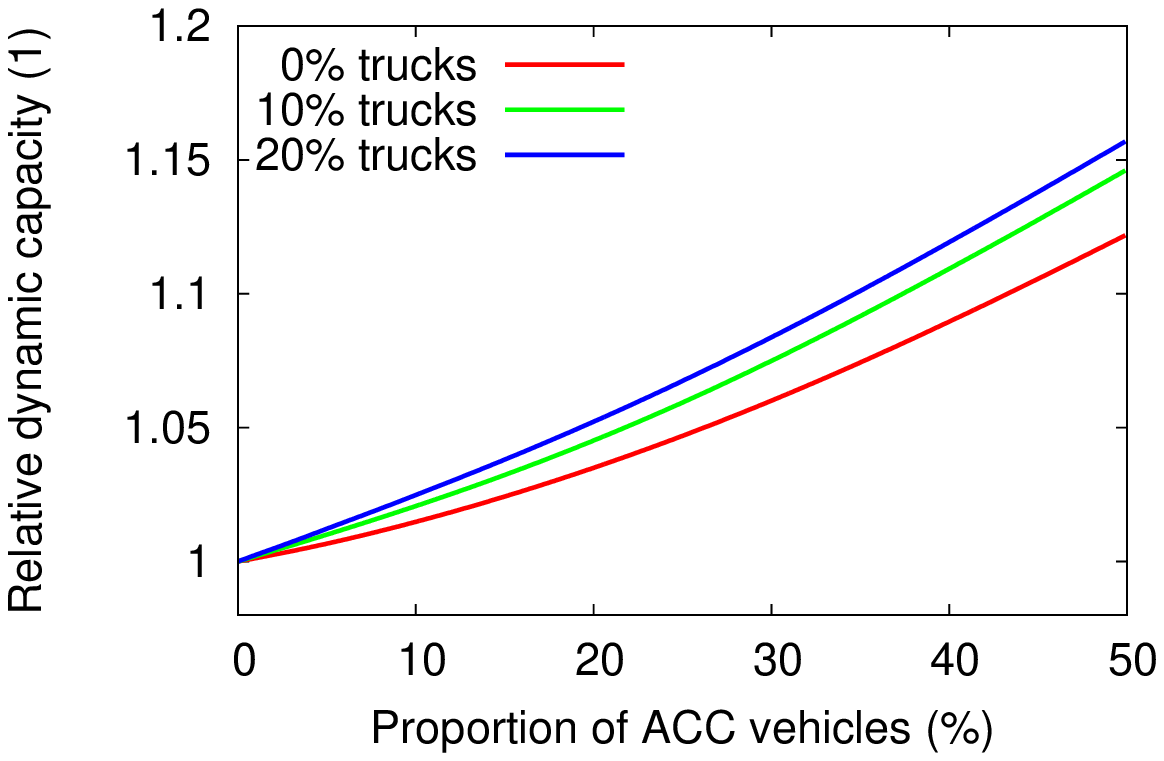}
\end{tabular}

 \caption{\label{fig:downstream}Dynamic capacity as a function of the
 percentage of ACC vehicles. The outflow from a traffic jam increases with a growing
 proportion  of ACC-equipped vehicles due to the changed driving strategy for the \gquote{downstream jam front} condition.}

\end{figure}

Figure~\ref{fig:downstream}(a) shows the resulting dynamic capacity for a
variable percentage of ACC vehicles in a scenario without trucks and
in a scenario with 10\% trucks. Single simulation runs are depicted by
symbols, while the average and the variation band were calculated
from the scattered simulation data via the kernel-based linear regression 
using a width of $\delta=0.1$ (cf.\ \ref{app:regression}). As intended by
the proposed traffic condition \gquote{downstream jam front}, the dynamic
capacity increases with growing ACC equipment rate.

Figure~\ref{fig:downstream}(b) compares the results for different truck
proportions by considering the \emph{relative increase
of the dynamic capacity} $q_\text{out} (\alpha) = Q_\text{out}(\alpha) / Q_\text{out}(0)$ with the ACC
equipment level $\alpha$. For $\alpha=50\%$,  the relative increase of $q_\text{out}$  is between
12\% and 16\% and therefore somewhat lower than the increase of the
maximum free capacity (cf.\ \S\ref{sec:sim}$\,$(\ref{sec:max_free}). Interestingly,
the dynamic capacity does not increase linearly as the measured
maximum free capacity displayed in figure~\ref{fig:free_vla}, but
faster. Consequently, the relative increase $q_\text{out}(\alpha)$
grows over-proportionally with higher ACC equipment rates $\alpha$. This can be understood
by an \gquote{obstruction effect} caused by slower
accelerating drivers (in particular, trucks) which hinder faster
(ACC) vehicles.

Furthermore, we can compare the maximum free capacity
$Q_\text{max}^\text{free}$ displayed in figure~\ref{fig:free_vla} with
the dynamic capacity $Q_\text{out}$, which is lower than
$Q_\text{max}^\text{free}$. The difference between both 
quantities is referred to as \emph{capacity drop}. Notice that the capacity drop is the crucial quantity determining
the performance (loss) of the freeway under congested conditions and accounts for the persistence of
traffic jams once the traffic flow has broken down. Realistic values
for the capacity drop are between 5\% and 20\%~\citep{Kerner-Rehb96,CasBer-99}. In our simulations, we found that the values of the relative capacity drop are typically between 5 and 15\%.

\subsubsection{Dynamic capacity for different driving strategy parameters}

\begin{figure}
\centering
\begin{tabular}{cc}
\includegraphics[width=0.45\textwidth]{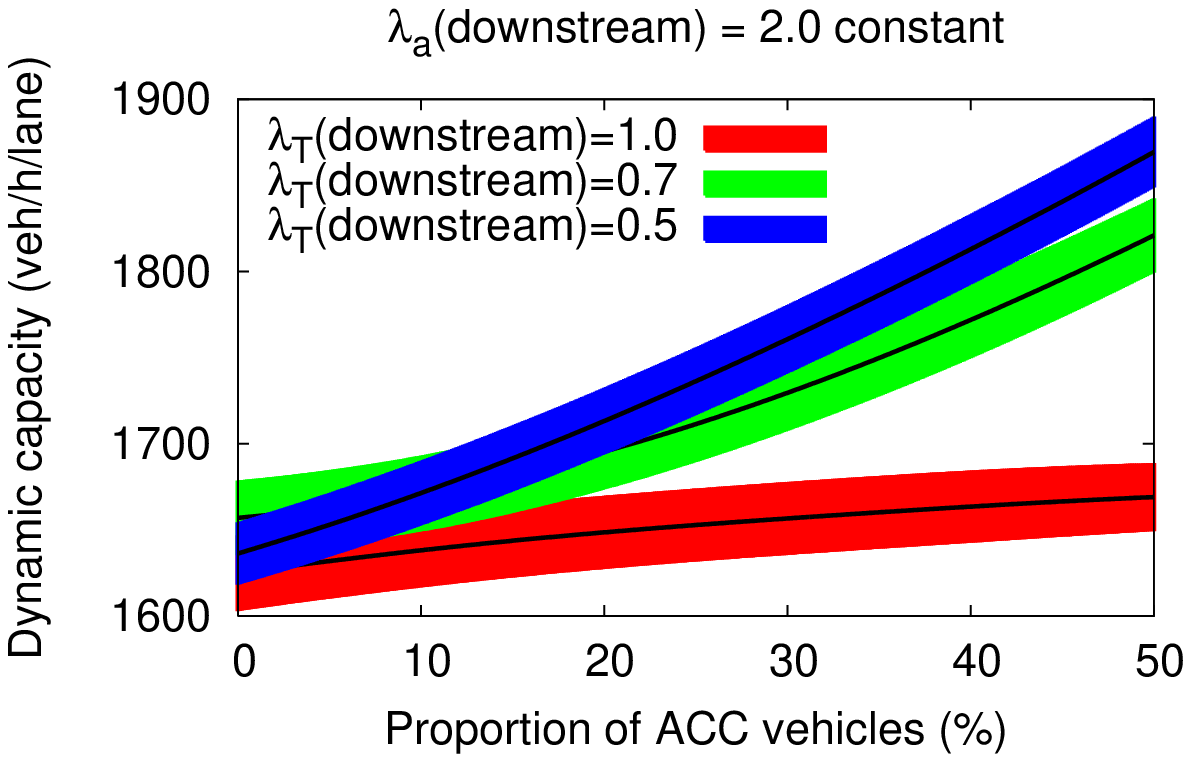} & 
\includegraphics[width=0.45\textwidth]{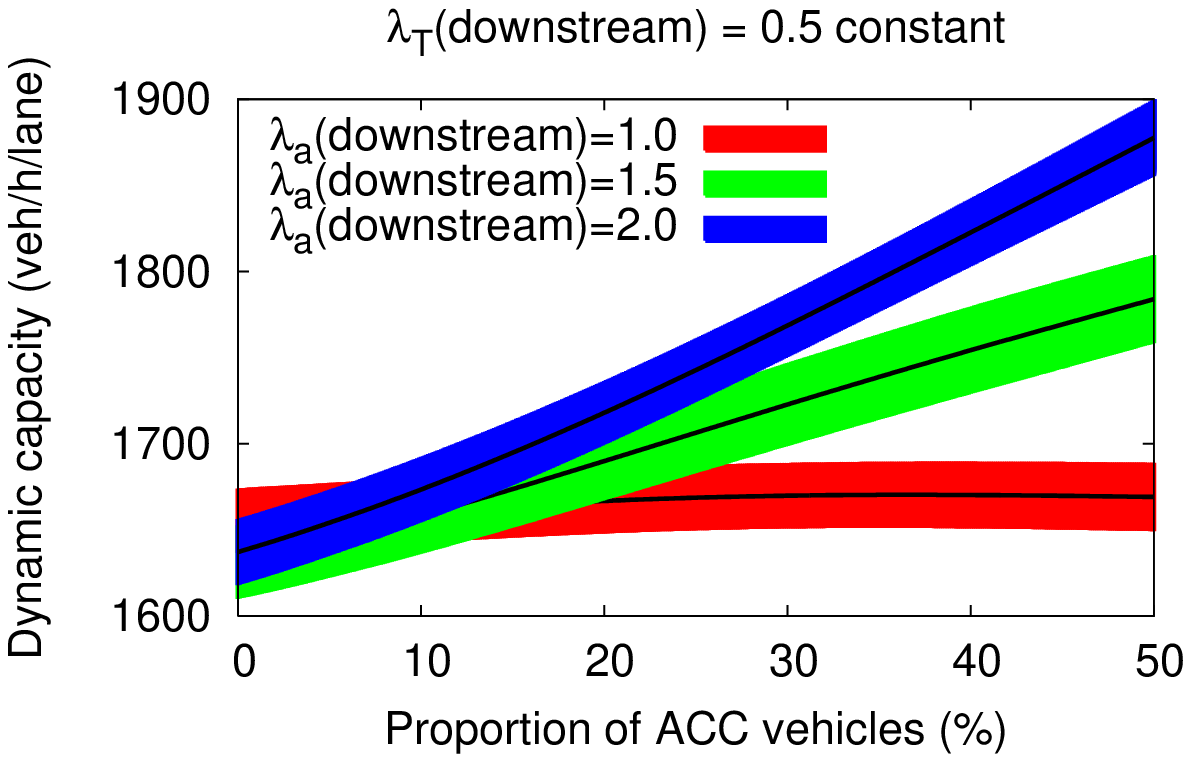} 
\end{tabular}

 \caption{\label{fig:mlc_aT}Dynamic capacity for various multiplication factors $\lambda$ used by the simulated ACC system in the \gquote{downstream jam front} traffic condition. The  results show that the dynamic
 capacity is increased in a relevant way only by adapting $a$ in
 combination with $T$.}

\end{figure}

The increase of the dynamic capacity is based on the proposed driving
strategy for the traffic condition \gquote{downstream jam front}. The
ACC-equipped vehicles increase their maximum acceleration $a$ in
combination with a decrease in the time gap $T$ (cf.\ table~\ref{tab:strategymatrix}). In figure~\ref{fig:mlc_aT}, simulation
results are shown for various multiplication factors $\lambda_a$ and
$\lambda_T$ for the traffic condition \gquote{downstream jam front} (using a
scenario with 10\% trucks). The default values
$\lambda_a^\text{down}=2$ and $\lambda_T^\text{down}=0.5$ correspond
to the results shown in figure~\ref{fig:downstream}. The simulation
results demonstrate that the dynamic capacity is only increased in a relevant
way by adapting $a$ \emph{and} $T$ simultaneously.

\section{\label{sec:disc}Discussion}

In view of an increasing number of vehicles equipped with adaptive
cruise control (ACC) systems in the future, it is important to address
questions about their impact and potentials on the collective traffic
dynamics. Until now, microscopic traffic models have mainly been used
to (approximately) describe the \emph{human driver}. For a realistic
description of the future mixed traffic, however, appropriate models
for both human driving and automated driving are needed. Car-following
models reacting only to the preceding vehicle have originally been
proposed to describe human drivers, but, from a formal point of view,
they describe more closely the dynamics of ACC-driven vehicles.
Thus, we have considered the simple
yet realistic Intelligent Driver Model (IDM)~\citep{Opus} as a
starting point for an adequate description of the driving behaviour on
a microscopic level. The IDM, however, is intended to describe traffic
dynamics in one lane only, and leads, e.g. in lane-changing
situations, to unrealistic driving behaviour when the actual gap is
significantly lower than the desired gap. For such situations, we have formulated an alternative
heuristic based on the assumption of constant accelerations in order
to prevent unnaturally strong braking reactions due to lane
changes. The proposed enhanced IDM combines the well-proven properties
of the original model with this \emph{constant-acceleration
heuristic} resulting in a more relaxed driving behaviour in situations
with are typically not considered to be critical by human drivers. The
new acceleration function~\eqref{accACC} depends (besides gap and
velocity) additionally on the acceleration of the leading vehicle.

Since the enhanced IDM is still a car-following model, we have called
it \gquote{ACC model}. In fact, it has already been implemented (with
some further confidential extensions) in real test cars  showing a realistic and natural driving dynamics~\citep{Kranke-Fisita2008,Kranke-VDI-2006-engl}. This consistency between automated driving and the \gquote{driving experience} of humans can be considered as a key account for the acceptance of and the confidence in ACC systems. Moreover, since the driver is obliged to override the ACC system at any point in time, an automated driving characteristics similar to those of human drivers is a safety-relevant premise. Hence, a \emph{realistic} car-following, i.e.,  an ACC model can be considered also as a appropriate description of the human driving although humans' perceptions and reactions fundamentally differ from those of ACC systems~\citep{HDM}.

The proposed ACC-model has been used to evaluate potentials of a vehicle-based approach for increasing traffic
performance in a mixed system consisting of human drivers and ACC
systems. While human drivers of cars and trucks have been modelled with
constant model parameters, the proposed adaptation of the driving
style according to the actual traffic conditions corresponds to
parameters that vary \emph{over time}. Note that the concrete meanings of the
IDM parameters allow for direct implementations of the considered
driving strategies.

By means of multi-lane traffic simulations, we have investigated the maximum
free flow before traffic breaks down (as crucial quantity in free traffic), and
the dynamic capacity (which determines as outflow from a traffic jam
the dynamics in congested traffic) by systematically increasing the
fraction of vehicles implementing multiple driving strategies. For the
maximum free flow, we found an approximately linear increase with a
sensitivity of about $0.3\%$ per 1\% ACC fraction. The dynamic
capacity shows a non-linear relationship with an increase of about
0.24\% per 1\% ACC fraction for small equipment rates. Both quantities
can be considered as \emph{generic} measures for the system
performance. These sensitivities multiply when considering other relevant measurements like travel
times resulting in large variations by factors of 2-4~\citep{Arne-ACC-TRC}. 
The presented results reveal the (positive) impact of the driving behaviour on traffic dynamics even when taking into consideration the
idealized assumptions and conditions in traffic simulations.

Finally, we note that the in-vehicle implementation of traffic-adaptive driving strategies and the detection of the proposed traffic conditions in real-time are investigated in ongoing research projects. Different technologies such as specifically attributed maps~\citep{Arne-ACC-TRC}, inter-vehicle communication~\citep{thiemann-IVC-PRE08,Kesting-IVC-Transactions09}, and vehicle-infrastructure integration~\citep{Kranke-Fisita2008} are considered at present.

\begin{acknowledgements}
The authors would like to thank Dr.~H.~Poppe and F.~Kranke for the excellent
collaboration and the Volkswagen AG for financial support within the
BMWi project AKTIV. 
\end{acknowledgements}

\appendix{\label{app:regression}Kernel-Based Linear Regression}
When dealing with simulations, one often varies a single model
parameter and plots over it a second quantity that results from the
related simulation run. Here, a statistical method is presented that combines the gradual change in the independent model
parameter $x$ with the estimation of the standard deviation in the resulting
fluctuating quantity $y$ without running multiple simulations for the
same parameter value $x$. 

Suppose we are fitting $n$ data points $\{x_i,y_i\},\, i=1,\ldots, n$
to a linear model with two parameters $a$ and $b$, $\hat{y}(x) = a + b x$.
A global measure for the goodness of fit is the sum of squared errors
$y_i - \hat{y}(x_i)$. The best-fitting
curve for the \emph{linear regression} can be obtained by the method of
least squares with respect to the fit parameters $a$ and $b$. The solution of the
system of linear equations is given by
\begin{eqnarray}\label{eq:fit_a}
b = b(\{x_i,y_i\}) &=& \frac{\avg{xy} - \avg{x}\avg{y}} {\avg{x^2} -
\avg{x}^2},\\ 
\label{eq:fit_b}
a = a(\{x_i,y_i\}) &=&  \frac{\avg{x^2}\avg{y} -
\avg{x}\avg{xy}}{\avg{x^2}-\avg{x}^2},
\end{eqnarray}
where the arithmetic average $\avg{z}$ of the measured data points
$\{z_i\}$ is defined by $\avg{z} := \frac{1}{n} \sum_{i=1}^n z_i$. Let us now generalize the linear regression by using a \emph{locally
weighted average}
\begin{equation}\label{eq:appendix_convolution}
\avg{z}(x) := \sum_{i=1}^n w(x - x_i)\, z_i,
\end{equation}
where the weights $w(x-x_i)$ are defined through a sufficiently
localized function $K$ with $w(x-x_i) = {K(x-x_i)} / \sum_j K(x-x_j)$.
As the expression~\eqref{eq:appendix_convolution} is evaluated locally
for any value $x$, the dependence on $x$ is transferred to the linear
fit para\-meters~\eqref{eq:fit_a} and~\eqref{eq:fit_b}.
For plotting $y$ against $x$, the special case of centering the
averages at $x'=x$ is relevant. With $a(x)=\avg{y}(x)-b(x)\,x$, we
obtain \begin{equation}\label{eq:weighted_lin_regr}
\hat{y}(x,x) = a(x)+b(x)\,x=\avg{y}(x).
\end{equation} 
Furthermore, the residual error $y_i - \hat{y}(x_i)$ is also weighted
by the discrete convolution~\eqref{eq:appendix_convolution}, resulting
in
\begin{equation}\label{eq:error_band}
\sigma^2(x) = \sum_{i=1}^n w_i(x)\left[ y_i - a(x) - b(x) x_i \right]^2.
\end{equation}
Notice that the \gquote{error band} $\sigma(x)$ describes the variations of the quantity $y$ on length scales smaller
than the width of the smoothing kernel $K$. For stochastic
simulations, this is simultaneously an estimate of the fluctuations
for given values of $x$. In this paper, we use a \emph{Gaussian
kernel} $K(x) =  \exp\left[-x^2 / (2\delta^2)\right]$ as weight
function. The width of the Gaussian kernel $\delta$ is the only parameter of the
kernel-based linear regression. 


\label{lastpage}
\end{document}